\documentclass[11pt]{article}
\usepackage{latexsym, amssymb, amsmath, amsthm}
\usepackage{graphicx,color,ulem}

\def \F {{\mathbb F}}

\def \T {{\rm Tr}}

\newtheorem{theorem}{Theorem}

\newtheorem{lemma}{Lemma}
\newtheorem{proposition}{Proposition}
\newtheorem{remark}{Remark}
\newtheorem{example}{Example}
\newtheorem{corollary}{Corollary}

\begin{document}

\begin{center}
{\Large \bf There are infinitely many bent functions for which the dual is not bent}\\[1em]
Ay\c ca \c Ce\c smelio\u glu$^{a}$, Wilfried Meidl$^b$, Alexander Pott$^c$ \\[.7em]
\end{center}
$^a$ \.Istanbul Kemerburgaz University, School of Arts and Sciences, Ba\u gc\i lar, 34217 \.Istanbul, Turkey. e-mail: ayca.cesmelioglu@kemerburgaz.edu.tr \\
$^b$ Johann Radon Institute for Computational and Applied Mathematics, Austrian Academy of Sciences, Altenbergerstrasse 69, 4040-Linz, Austria. e-mail: meidlwilfried@gmail.com \\
$^c$ Otto-von-Guericke-University, Faculty of Mathematics, 39106 Magdeburg, Germany. e-mail: alexander.pott@ovgu.de\\

\begin{abstract}
Bent functions can be classified into regular bent functions, weakly regular but not regular bent functions,
and non-weakly regular bent functions. Regular and weakly regular bent functions always appear in pairs since their duals are also
bent functions. In general this does not apply to non-weaky regular bent functions. However, the first known construction of non-weakly 
regular bent functions by Ce\c smelio\u glu et {\it al.}, 2012, yields bent functions for which the dual is also bent. In this paper 
the first construction
of non-weakly regular bent functions for which the dual is not bent is presented. We call such functions non-dual-bent functions. 
Until now, only sporadic examples found via computer search were known. We then show that with the 
direct sum of bent functions and with the construction by Ce\c smelio\u glu et {\it al.} one can obtain infinitely many non-dual-bent 
functions once one example of a non-dual-bent function is known. 
\end{abstract}

\section{Introduction}

For a prime $p$, let $f$ be a function from an $n$-dimensional vector space $V_n$ over $\F_p$ to $\F_p$.
The {\it Walsh transform} of $f$ is the complex valued function
\[ \widehat{f}(b) = \sum_{x\in V_n}\epsilon_p^{f(x)-\langle b,x\rangle}, \quad \epsilon_p = e^{2\pi i/p},  \]
where $\langle b,x\rangle$ is a (nondegenerate) inner product in $V_n$. The function $f$ is called a 
{\it bent function} if $|\widehat{f}(b)| = p^{n/2}$ for all $b\in V_n$.
For Boolean bent functions we have $\widehat{f}(b) = (-1)^{f^*(b)}2^{n/2}$ for a Boolean function $f^*$, 
called the dual of $f$. When $p$ is odd, then a bent function $f$ satisfies (cf. \cite{hk})
\begin{equation}
\label{(2)} 
\widehat{f}(b) =
\left\{\begin{array}{r@{\quad:\quad}l}
\pm \epsilon_p^{f^*(b)}p^{n/2} & p^n \equiv 1\bmod 4; \\
\pm i\epsilon_p^{f^*(b)}p^{n/2} & p^n \equiv 3\bmod 4,
\end{array}\right.
\end{equation}
for a function $f^*$ from $V_n$ to $\F_p$. Accordingly $f$ is called 
{\it regular} if $p^{-n/2}\widehat{f}(b) =\epsilon_p^{f^*(b)}$ for all $b \in V_n$,
which for a Boolean bent function always holds.
If $p^{-n/2}\widehat{f}(b) =\zeta\ \epsilon_p^{f^*(b)}$ for some $\zeta\in\{\pm 1,\pm i\}$, independent from $b$,
we call $f$ {\it weakly regular}, otherwise $f$ is called {\it non-weakly regular}.
Note that regular implies weakly regular. 

Weakly regular bent functions $f$ always appear in pairs, as also the dual $f^*$ of $f$ is bent.
We restate here the argument in \cite{hk}, see also \cite{cmp}: \\
For $y\in V_n$ we get
\begin{equation}
\label{poisson}
\sum_{b\in V_n}\epsilon_p^{\langle b,y\rangle}\widehat{f}(b) =  
\sum_{b\in V_n}\epsilon_p^{\langle b,y\rangle}\sum_{x\in V_n}\epsilon_p^{f(x)-\langle b,x\rangle} 
=\sum_{x\in V_n}\epsilon_p^{f(x)}\sum_{b\in V_n}\epsilon_p^{\langle b,(y-x)\rangle} 
= p^n\epsilon_p^{f(y)},
\end{equation}
a special case of {\it Poisson Summation Formula}.
We now use that $f$ is weakly regular, hence $\widehat{f}(b) = \zeta p^{n/2}\epsilon_p^{f^*(b)}$, with 
$\zeta$ fixed, independent from $b$. Then
\[ p^n\epsilon_p^{f(y)} = \zeta
p^{n/2}\sum_{b\in V_n}\epsilon_p^{f^*(b)+\langle b,y\rangle} = \zeta p^{n/2}\widehat{f^*}(-y). \]
Consequently
\begin{equation}
\label{thedual}
\widehat{f^*}(-y) = \zeta^{-1}p^{n/2}\epsilon_p^{f(y)}
\end{equation}
and therefore $f^*$ is weakly regular bent. 

All classical constructions of bent functions yield weakly regular bent functions. The first sporadic examples of
non-weakly regular bent functions, all in characteristic $3$ and found by computer search, appeared in \cite{hk}, \cite{hk1} and \cite{hk2}. 
In \cite{tyz}, it was observed that one can obtain more examples in dimension $m+n$ with the direct sum $F(x,y) = f(x)+g(y)$ 
if one chooses for $f$ a (weakly) regular bent function in dimension $m$ and for $g$ a non-weakly regular bent function
in dimension $n$. The first construction of infinite classes of non-weakly regular bent functions was given in \cite{agw},
and further analysed in \cite{aw,aw1,cmp,cmp1}. The results indicate that though the ``obvious'' constructions yield (weakly) regular bent 
functions, being non-weakly regular is not at all an exceptional property for a bent function (in odd characteristic).

In \cite{cmp} it was observed that the construction of non-weakly regular bent functions in \cite{agw}, which uses previously 
known bent functions in dimension $n$ to obtain one bent function in dimension $n+2$, yields bent functions for which the dual
is also a bent function. On the other hand, some of the found sporadic examples of non-weakly regular bent functions do not
have a bent dual:
\begin{enumerate}
\item $g_1:\F_{3^6}\rightarrow \F_3$ with $g_1(x) = {\rm Tr}_6(\xi^7x^{98})$, where
$\xi$ is a primitive element of $\F_{3^6}$, see \cite{hk},
\item $g_2:\F_{3^4}\rightarrow \F_3$ with $g_2(x) = {\rm Tr}_4(a_0x^{22}+x^4)$, where
$a_0\in\{\pm\xi^{10},\pm\xi^{30}\}$ and $\xi$ is a primitive element of $\F_{3^4}$, 
see \cite{hk1},
\item $g_3:\F_{3^6}\rightarrow \F_3$ with $g_5(x) = {\rm Tr}_6(\xi^7 x^{14}+ \xi^{35}x^{70})$, where
$\xi$ is a primitive element of $\F_{3^6}$, see \cite{hk2}.
\end{enumerate}
Following this observation, in \cite{cmp} a new concept of bent functions was introduced:
A bent function $f$ is called a {\it dual-bent function} if the dual function $f^*$ defined as in $(\ref{(2)})$
is also bent. Otherwise we call $f$ a {\it non-dual-bent function}. 
Clearly every weakly regular bent function is a dual-bent function, but the converse does not hold. 

For all types of dual-bent functions, regular, weakly regular but not regular, and non-weakly regular but dual-bent, we
know constructions. What is missing, is a theoretical 
construction of non-dual-bent functions, i.e. a construction of bent functions
for which the dual is not a bent function.

The objective of this paper is to close this gap, presenting the first construction of bent functions which yields
non-dual-bent functions. In Section \ref{sec2} we present a construction of bent functions which can be seen as an
extension of the direct sum. In Section \ref{sec3} we show that this construction in general yields non-dual-bent
functions, and we give some examples of non-dual-bent functions. In Section \ref{sec4} we show that with the direct 
sum of bent functions and with the construction in \cite{agw} one can obtain infinitely
many non-dual-bent functions once one example of a non-dual-bent function is constructed.

\section{A semi-direct sum of bent functions}
\label{sec2}

A simple construction of a new bent function from two given bent functions is the {\it direct sum} of a bent function $f$ from
$V_m$ to $\F_p$ and a bent function $g$ from $V_n$ to $\F_p$, which is the function $F:V_m\times V_n\rightarrow\F_p$ defined as 
$F(x,y) = f(x)+g(y)$. It is straightforward to determine that
\[ \widehat{F}(a,b) = \widehat{f}(a)\widehat{g}(b). \] 
In this section we present an extension of this secondary construction which we may call the {\it semi-direct sum} of two
bent functions $f$ and $g$. We will employ this semi-direct sum in the next section to provide the first construction of non-dual-bent functions.
\begin{theorem}
\label{sds}
Let $f:V_m\rightarrow\F_p$ and $g:V_n\rightarrow\F_p$ be bent, and let $h$ be a function from $V_m$ to $V_n$. 
The function $F:V_m\times V_n\rightarrow\F_p$ defined as
\begin{equation}
\label{Fxy}
F(x,y) = f(x) + g(y+h(x)) 
\end{equation}
is bent if and only if for all $b\in V_n$ the function $G_b:V_m\rightarrow\F_p$
\[ G_b(x) = f(x) + \langle b,h(x)\rangle \]
is a bent function. The dual $F^*$ of $F$ is then
\[ F^*(x,y) = G_y^*(x)+g^*(y). \]
\end{theorem}
{\it Proof.}
For $a\in V_m$ and $b\in V_n$ we have
\begin{eqnarray}
\label{whF}
\nonumber
\widehat{F}(a,b) & = & \sum_{x\in V_m, y\in V_n}\epsilon_p^{f(x) + g(y+h(x)) - \langle a,x\rangle - \langle b,y\rangle} \\ \nonumber
& = & \sum_{x\in V_m}\epsilon_p^{f(x) - \langle a,x\rangle}\sum_{y\in V_n}\epsilon_p^{g(y) - \langle b,y-h(x)\rangle} \\ \nonumber
& = & \sum_{x\in V_m}\epsilon_p^{f(x) + \langle b,h(x)\rangle - \langle a,x\rangle}\sum_{y\in V_n}\epsilon_p^{g(y) - \langle b,y\rangle} \\
& = & \widehat{G_b}(a)\widehat{g}(b).
\end{eqnarray}
Since $g$ is bent, i.e. $\widehat{g}(b) = \zeta p^{n/2}\epsilon_p^{g^*(b)}$ for some $\zeta\in\{\pm 1,\pm i\}$ (which may depend on $b$),
the function $F$ is bent if and only if $|\widehat{G_b}(a)| = p^{m/2}$ for all $a\in V_m$ and $b\in V_n$, or equivalently $G_b$ is bent
for all $b\in V_n$. Then
\[ \widehat{F}(a,b) = \zeta p^{(m+n)/2}\epsilon_p^{G_b^*(a)+g^*(b)} \]
for some $\zeta\in\{\pm 1,\pm i\}$ (which may depend on $a$ and $b$), and the formula for the dual $F^*$ follows.\hfill$\Box$
\begin{remark}
If $h$ is the zero function, then the condition in Theorem \ref{sds} trivially holds and the semi-direct sum reduces to the direct sum.
\end{remark}
\begin{remark}
In \cite{c}, Carlet presented the special case of the construction in Theorem \ref{sds} where $p=2$ and $g$ is the quadratic 
Maiorana-McFarland bent function $g(x) = x_1x_2+x_3x_4+\cdots+x_{n-1}x_n$ from $\F_2^n$ to $\F_2$, $n$ even. The function 
$F:V_m\times\F_2^n\rightarrow\F_2$ is then of the form
\begin{equation}
\label{F} 
F(x,y_1,\ldots,y_n) = f(x) + \sum_{i=1}^{n/2}(y_{2i-1}+h_{2i-1}(x))(y_{2i}+h_{2i}(x))
\end{equation}
for some functions $h_1,\ldots,h_n$ from $V_m$ to $\F_2$.
\end{remark}

%
%
%

\section{Bent functions for which the dual is not bent}
\label{sec3}

We first present some examples of bent functions from $V_m\times\F_p^n$ to $\F_p$ obtained with the semi-direct sum. Note that to satisfy the conditions in Theorem \ref{sds} 
we need a bent function $f:V_m\rightarrow\F_p$ and a function $h(x) = (h_1(x),h_2(x),\ldots,h_n(x))$ from $V_m$ to $\F_p^n$ such that for all $\Lambda = (\lambda_1,\ldots,\lambda_n)\in \F_p^n$
the function
\[ G_\Lambda(x) = f(x) + \Lambda\cdot h(x) = f(x)+\lambda_1h_1(x) + \lambda_2h_2(x) + \cdots + \lambda_nh_n(x) \]
is bent. To generate such functions $f,h_1,\ldots,h_n$ from $V_m$ to $\F_p$ we will employ vectorial bent functions from $\F_{p^m}$ to $\F_p$.
Hence in this section $V_m$ will be identified with the finite field $\F_{p^m}$.
Examples for vectorial bent functions are quadratic PN-functions of which the simplest are quadratic monomials, or the Coulter-Matthews functions for $p=3$.
\begin{lemma} [Lemma 2 and Corollary 3 in \cite{hk}]
\label{hklem}
Let $m$ and $0\le k\le m$ be integers such that $m/\gcd(m,k)$ is odd. For a nonzero $\alpha\in\F_{p^m}$ let $f_\alpha$ be the function
$f_\alpha(x) = \T_m(\alpha x^{p^k+1})$ from $\F_{p^m}$ to $\F_p$. Then
\[ \widehat{f_\alpha}(u) = \left\{\begin{array}{l@{\quad:\quad}l}
\eta(\alpha)(-1)^{m-1}p^{m/2}\epsilon_p^{f_\alpha^*(u)} & p\equiv 1 (\bmod 4) \\
\eta(\alpha)(-1)^{m-1}i^mp^{m/2}\epsilon_p^{f_\alpha^*(u)} & p\equiv 3 (\bmod 4),
\end{array}\right. \]
where $\eta(\alpha)$ denotes the quadratic character of $\alpha$ in $\F_{p^m}$.
\end{lemma}
\begin{lemma}[see  Lemma 2 in \cite{fl} and Proposition 2 in \cite{cmp1}]
\label{CM}
Let $m,k$ be positive integers such that $\gcd(2m,k)=1$.
For each $\alpha \in \F_{3^m}^*$, the Walsh transform $\widehat{f_{\alpha}}$ of the weakly regular bent function 
$f_{\alpha}(x)=\T_m(\alpha x^{\frac{3^k+1}{2}})$ satisfies
\[\widehat{f_{\alpha}}(u)=\eta(\alpha)(-1)^{m-1}i^m3^{m/2}\epsilon_3^{f_\alpha^*(u)}, \]
where $\eta(\alpha)$ denotes the quadratic character of $\alpha$ in $\F_{3^m}$.
\end{lemma}
\begin{corollary}
\label{cor}
For integers $m$ and $n$, $2\le n<m$, and let $\alpha_0,\alpha_1,\ldots,\alpha_n\in\F_{p^m}$ be linearly independent over $\F_p$,
and let $g$ be a weakly regular bent function from $\F_p^n$ to $\F_p$. Let 
\begin{itemize}
\item[-] $G(x) = x^{p^k+1}$ for some integer $0\le k\le m$ such that $m/\gcd(m,k)$ is odd, and $f_{\alpha_j}(x) = \T_m(\alpha_jG(x))$,
$0\le j \le n$, or
\item[-] $p=3$ and $G(x) = x^{\frac{3^k+1}{2}}$ for an integer $k$ such that $\gcd(2m,k)=1$ and $f_{\alpha_j}(x) = \T_m(\alpha_jG(x))$,
$0\le j \le n$.
\end{itemize}
Then $F:\F_{p^m}\times\F_p^n\rightarrow\F_p$
\[ F(x,y_1,\ldots,y_n) = f_{\alpha_0}(x) + g(y_1+f_{\alpha_1}(x),y_2+f_{\alpha_2}(x),\ldots,y_n+f_{\alpha_n}(x)). \]
is a bent function, which in general is non-weakly regular.
\end{corollary}
{\it Proof.}
Clearly, $F$ is the bent function $(\ref{Fxy})$ for $f=f_{\alpha_0}$ and $h_j = f_{\alpha_j}$, $1\le j\le n$.
Note that $G_{\lambda_1,\ldots,\lambda_n}(x) = \T_m((\alpha_0+\sum_{j=1}^n\lambda_j\alpha_j)G(x))$ is bent since the elements $\alpha_j$ 
are chosen to be linearly independent, hence $\alpha_0+\sum_{j=1}^n\lambda_j\alpha_j \ne 0$.
To see that $F$ is non-weakly regular, we choose $\lambda_1,\ldots,\lambda_n$ and $\bar{\lambda}_1,\ldots,\bar{\lambda}_n$
such that $\Lambda = \alpha_0+\sum_{j=1}^n\lambda_j\alpha_j$ is a square and $\bar{\Lambda} = \alpha_0+\sum_{j=1}^n\bar{\lambda}_j\alpha_j$ 
is a non-square in $\F_{p^m}$. Clearly in general such $\Lambda$, $\bar{\Lambda}$ exist.
With Lemma \ref{hklem} respectively Lemma \ref{CM}, and the assumption that $\eta(\Lambda) \ne \eta(\bar{\Lambda})$,
the non-weak regularity of $F$ follows from $(\ref{whF})$ together with the weak regularity of $g$. \hfill$\Box$\\[.5em]
In the remainder of this section we show that in general the construction in Theorem \ref{sds} yields bent functions for which 
the dual is not bent. Our functions are the first theoretically constructed non-dual-bent functions. We follow the approach of Corollary \ref{cor}
where we employ vectorial bent functions for our construction. For simplicity we choose $n=2$ and $G(x) = x^2$ and $g(y_1,y_2) = y_1y_2$. Then
\[ F(x,y_1,y_2) = f(x)+(y_1+h_1(x))(y_2+h_2(x)) \]
where $f(x) = \T_m(x^2)$, $h_1(x) = \T_m(\alpha x^2)$, $h_2(x) = \T_m(\beta x^2)$ and $1,\alpha,\beta$ are linearly independent over $\F_p$
(we take $\alpha_0 = 1$, $\alpha_1 = \alpha$, $\alpha_2 = \beta$). 
As an application of Theorem \ref{sds} we obtain the subsequent corollary.
\begin{corollary}
\label{nd-cor}
Let $1,\alpha, \beta \in \F_{p^m}$ be linearly independent over $\F_p$. If
\begin{equation}
\label{condition}
|\sum_{y_1,y_2\in\F_p}\eta(1+y_1\alpha+y_2\beta)\epsilon_p^{-y_1y_2}| \ne p, 
\end{equation}
then the function $F:\F_{p^m}\times\F_p^2$
\begin{equation}
\label{bentform}
F(x,y_1,y_2) = \T_m(x^2)+(y_1+\T_m(\alpha x^2))(y_2+\T_m(\beta x^2))
\end{equation}
is a non-dual-bent function.
\end{corollary}
{\it Proof.}
Observing that $g^*(y_1,y_2) = -y_1y_2$, the dual of $F$ is
\[ F^*(x,y_1,y_2) = G^*_{y_1,y_2}(x)-y_1y_2 \]
where $G^*_{y_1,y_2}(x)$ is the dual of $G_{y_1,y_2}(x) = \T_m((1+y_1\alpha+y_2\beta)x^2)$.
By \cite[Corollary 3]{hk}, 
\[ G^*_{y_1,y_2}(x) = -\T_m(\frac{x^2}{4(1+y_1\alpha+y_2\beta)}). \] 
Furthermore,
\[ G^{**}_{y_1,y_2}(x) = G_{y_1,y_2}(-x) = \T_m((1+y_1\alpha+y_2\beta)x^2) = G_{y_1,y_2}(x). \] 
We determine the Walsh coefficient of $F^*$ at $(0,0,0)$:
\begin{eqnarray*}
\widehat{F^*}(0,0,0) & = & \sum_{x\in\F_{p^m}\atop y_1,y_2\in\F_p}\epsilon_p^{G^*_{y_1,y_2}(x)-y_1y_2} = 
\sum_{y_1,y_2\in\F_p}\epsilon_p^{-y_1y_2}\sum_{x\in\F_{p^m}}\epsilon_p^{G^*_{y_1,y_2}(x)} \\
& = & \sum_{y_1,y_2\in\F_p}\epsilon_p^{-y_1y_2}\widehat{G^*_{y_1,y_2}}(0)
= \zeta p^{m/2}\sum_{y_1,y_2\in\F_p}\epsilon_p^{-y_1y_2}\eta(1+y_1\alpha+y_2\beta)\epsilon_p^{G^{**}_{y_1,y_2}(0)} \\
& = & \zeta p^{m/2}\sum_{y_1,y_2\in\F_p}\eta(1+y_1\alpha+y_2\beta)\epsilon_p^{-y_1y_2},
\end{eqnarray*}
where $\zeta\in\{\pm 1, \pm i\}$ only depends on $p$ and $m$, see Lemma \ref{hklem}. As a consequence, if
\[ |\sum_{y_1,y_2\in\F_p}\eta(1+y_1\alpha+y_2\beta)\epsilon_p^{-y_1y_2}| \ne p, \]
then $F^*$ is not bent. \hfill$\Box$\\[.5em]
Condition $(\ref{condition})$ combines the additive and the multiplicative structure of the finite field
and is therefore not easy to analyse. If all values for $1+\lambda_1\alpha+\lambda_2\beta$, $\lambda_1,\lambda_2\in\F_p$,
have the same quadratic character, then $F$ is weakly regular, hence a dual-bent function. As obvious, in this case
the character sum in $(\ref{condition})$ has in fact absolute value $p$. Clearly with a random choice of $\alpha, \beta$ 
this is quite unlikely, and one also would expect a chaotic behaviour of the character sum in $(\ref{condition})$.
In particular it seems that its absolute value is rarely $p$.
Below are some examples of non-dual-bent functions obtained with Corollary \ref{nd-cor} for $p=3$ 
%
and for $p=5$. 
%
%
%
%
\begin{example}
\label{ex1}
Let $p=3, m=3$, and let $w$ be a root of the irreducible polynomial $g(x)=x^3+x^2+2 \in \F_3[x]$. For both choices
\begin{itemize}
\item[(i)] $\alpha = w$, $\beta = w^2+1$,
\item[(ii)] $\alpha = 2w+1$, $\beta = w^2$
\end{itemize}
the character sum in $(\ref{condition})$ has absolute value $\sqrt{3}$. Hence in both cases the dual of the bent function
$F(x,y_1,y_2) = \T_3(x^2)+(y_1+\T_3(\alpha x^2))(y_2+\T_3(\beta x^2))$ from $\F_{3^3}\times\F_3^2\rightarrow\F_3$ is not a bent function.
\end{example}
\begin{example}
Let $w\in\F_{3^4}$ be a root of the irreducible polynomial $g(x)=x^4+x^3+2\in\F_3[x]$. For $\alpha = w$ and $\beta = w^2$ we have
$|\sum_{y_1,y_2\in\F_3}\eta(1+y_1\alpha+y_2\beta)\epsilon_p^{-y_1y_2}| = |1-2\sqrt{3}i| = \sqrt{13} \ne 3$. Hence the dual of the
bent function $F(x,y_1,y_2)=T_3(x^2)+(y_1+\T_3(wx^2))(y_2+\T_3(w^2x^2))$ from $\F_{3^4}\times\F_3^2$ to $\F_3$ is not bent.
\end{example}
\begin{example}
Let $w\in\F_{5^3}$ be a root of the irreducible polynomial $g(x)=x^3+x+1$, let $\alpha = \omega$, $\beta = \omega^2$,
and let $F$ be the bent function from $\F_{5^3}\times\F_5^2$ to $\F_5$
given by $F(x,y_1,y_2)=T_5(x^2)+(y_1+\T_5(wx^2))(y_2+\T_5(w^2x^2))$. Since $|\sum_{y_1,y_2\in\F_5}\eta(1+y_1\alpha+y_2\beta)\epsilon_p^{-y_1y_2}| =
|4\epsilon_5^4-4\epsilon_5+1| \ne 5$, the dual of $F$ is not a bent function. We remark that all previously known sporadic examples of non-dual-bent 
functions are in characteristic $3$. With our 
choice of $\alpha$ and $\beta$, checking the condition in Corollary \ref{nd-cor} without difficulty
we constructed a non-dual-bent function in characteristic $5$.
\end{example}
We add an example of a bent function of the form $(\ref{bentform})$ for which the absolute value of the character sum in Corollary \ref{nd-cor}
equals $p$.
\begin{example}
Let $w\in\F_{3^3}$ be as in Example \ref{ex1}, and choose $\alpha = w$ and $\beta = w^2$. 
In this case we have $|\sum_{y_1,y_2\in\F_3}\eta(1+y_1\alpha+y_2\beta)\epsilon_p^{-y_1y_2}| = 3$. Hence for the corresponding bent function $F$
we have $|\widehat{F}(0)| = 3^{5/2}$, and $F$ may or may not have a bent dual. Using Magma we confirmed that the dual of $F$ is again not a bent 
function.
\end{example}

\section{Recursively constructing non-dual-bent functions}
\label{sec4}

In this section we show that once a non-dual-bent function is constructed, one can recursively obtain infinitely many with the
direct sum and the construction in \cite{agw}. We emphasize that these secondary constructions of bent functions cannot provide
non-dual-bent functions if one does not use a bent function as building block
which is already non-dual.

Recall that for two functions $f:V_m\rightarrow\F_p$ and $g:V_n\rightarrow\F_p$ the direct sum $F:V_m\times V_n\rightarrow\F_p$
is defined as $F(x,y) = f(x)+g(y)$. As easily seen
\[ \widehat{F}(a,b) = \widehat{f}(a)\widehat{g}(b). \]
In particular if $f$ and $g$ are bent, then $F$ is bent, and
\[ \widehat{F}(a,b) = \zeta_{a,b}p^{\frac{m+n}{2}}\epsilon_p^{f^*(a)+g^*(b)} \]
for some $\zeta_{a,b}\in\{\pm 1, \pm i\}$. Hence the dual of $F$ is 
$F^*(x,y) = f^*(x)+g^*(y)$.
\begin{theorem}
The direct sum of a dual-bent function and a non-dual-bent function is a non-dual-bent function.
\end{theorem}
{\it Proof.}
Suppose that $f^*$ is bent but $g^*$ is not, hence $|\widehat{g^*}(b)| = A \ne p^{n/2}$ for some $b\in V_n$. Then
\[ |\widehat{F^*}(a,b)| = p^{m/2}A \ne p^{(m+n)/2} \]
for all $a\in V_m$, which finishes the proof.
\hfill$\Box$ \\[.5em]
Now we consider the construction introduced in \cite{agw} and further investigated in \cite{aw,aw1,cmp,cmp1}, which combines $p$ bent 
functions from $V_n$ to $\F_p$ to one bent function in dimension $n+2$. We follow the notation in \cite{cmp} and use the multivariate
representation, i.e. we represent $V_n$ as $\F_p^n$.
\begin{proposition}
\label{T2} 
For $j=0,\ldots,p-1$ let $f_j$ be functions from $\F_p^n$ to $\F_p$. The function $F:\F_p^{n+2}\rightarrow\F_p$
defined as 
\[ F(x,x_{n+1},y) = f_y(x)+x_{n+1}y \]
is bent if and only if for all $0\le j\le p-1$ the function $f_j$ is bent.
\end{proposition}
%
%
{\it Proof.} 
By definition, $F$ is bent if for all $a \in \F_p^n$, $b,c\in\F_p$ the Walsh transform $\widehat{F}(a,b,c)$
has absolute value $p^{(n+2)/2}$. For $a \in \F_p^n$, $b,c\in\F_p$ we have
\begin{eqnarray*}
\widehat{F}(a,b,c) & = & \sum_{x\in\F_p^n\atop x_{n+1},y\in\F_p} \epsilon_p^{f_y(x)+x_{n+1}y-a\cdot x -bx_{n+1}-cy} \\
& = & \sum_{x\in\F_p^n\atop y\in\F_p}\epsilon_p^{f_y(x)-a\cdot x-cy}\sum_{x_{n+1}\in\F_p}\epsilon_p^{x_{n+1}(y-b)}
= p\epsilon_p^{-bc}\sum_{x\in\F_p^n}\epsilon_p^{f_b(x)-a\cdot x} \\
& = & p\epsilon_p^{-bc}\widehat{f_b}(a).
\end{eqnarray*}
Consequently $|\widehat{F}(a,b,c)| = p^{(n+2)/2}$ for all $a \in \F_p^n$, $b,c\in\F_p$ if and only if $|\widehat{f_b}(a)| = p^{n/2}$
for all $a \in \F_p^n$, $b\in\F_p$, which applies if and only if $f_b$ is bent for all $0\le b\le p-1$.
\hfill$\Box$\\[.5em]
We observe that if $f_b$, $0\le b\le p-1$, is bent, then $\widehat{F}(a,b,c) = p^{\frac{n+2}{2}}\zeta\epsilon_p^{f^*_b(a)-bc}$. 
Consequently, as also observed in \cite{cmp}, the dual of the bent function $F$
\begin{equation}
\label{F3} 
F^*(x,x_{n+1},y) = f^*_{x_{n+1}}(x)-x_{n+1}y,
\end{equation}
is obtained with the same construction method from $f_j^*$, $0\le j\le p-1$, (the roles of the variables $x_{n+1}$ and $y$ are interchanged).
With those observations and Proposition \ref{T2} we get the following theorem.
\begin{theorem}
For $j=0,\ldots,p-1$ let $f_j$ be bent functions from $\F_p^n$ to $\F_p$. The bent function $F:\F_p^{n+2}\rightarrow\F_p$
defined as 
\[ F(x,x_{n+1},y) = f_y(x)+x_{n+1}y \]
is dual-bent if and only if for all $0\le j\le p-1$ the function $f_j$ is dual-bent.
\end{theorem}

\section{Concluding remarks}

In the literature many constructions and explicit representations of bent functions, also in odd characteristic,
can be found. Almost all of them describe (weakly) regular bent functions. In \cite{agw} the first construction of
infinite classes of non-weakly regular bent functions has been presented. This construction combines regular and
weakly regular (but not regular) bent functions in dimension $n$, of which many infinite classes are known, to
one non-weakly regular bent function in $n+2$ variables. As observed in \cite{cmp}, the resulting bent functions
are dual-bent, a property which non-weakly regular bent functions do not
necessarily have. In this article the first theoretical
construction of non-dual-bent functions is presented. Moreover we show that with the direct sum of bent functions 
and with the construction in \cite{agw}, recursively one can obtain infinitely many non-dual-bent functions once 
one example of a non-dual-bent function is known. Our results indicate that being non-dual-bent is not an 
exceptional property for a bent function. \\[1em]

\noindent
{\bf Acknowledgement.}
The second author is supported by the Austrian Science Fund (FWF) Project no. M 1767-N26.

\end{document}